\documentclass[a4paper]{article}%
\usepackage{amsmath}
\usepackage{amsfonts}
\usepackage{authblk}
\usepackage{amssymb}
\usepackage{graphicx}
\usepackage[top=2cm,bottom=2cm,right=2cm,left=2cm]{geometry}%
\usepackage{hyperref,natbib,bm,url}
\usepackage{multirow}

\author[1,2]{Andr\'{e} Mas\thanks{Corresponding author: \texttt{andre.mas@umontpellier.fr}}}
\author[1]{Cl\'{e}ment Carr\'{e}}
 
\affil[1]{Bionomeex, Montpellier, France }
\affil[2]{IMAG, CNRS, Univ. Montpellier, France}

\begin{document}

\title{Prediction of Hilbertian autoregressive processes : a Recurrent Neural Network approach \protect}

\date{}
\maketitle

%\affiliation{BionomeeX, Montpellier, France \thanksmark{m1} and IMAG, Univ. Montpellier, CNRS, Montpellier, France\thanksmark{m2}}

\begin{abstract}
The autoregressive Hilbertian model (ARH) was introduced in the early 90's by Denis Bosq. It was the subject of a vast literature and gave birth to numerous extensions. The model generalizes the classical multidimensional autoregressive model, widely used in Time Series Analysis. It was successfully applied in numerous fields such as finance, industry, biology. We propose here to compare the classical prediction methodology based on the estimation of the autocorrelation operator with a neural network learning approach. The latter is based on a popular version of Recurrent Neural Networks : the Long Short Term Memory networks. The comparison is carried out through simulations and real datasets.
\end{abstract}

%\begin{keyword}[class=AMS]
%\kwd[Primary ]{60K35}
%\kwd{60K35}
%\kwd[; secondary ]{60K35}
%\end{keyword}

%\begin{keyword}
%\kwd{Functional Data}
%\kwd{ARH process}
%\kwd{LSTM networks}
%\kwd{Prediction}
%\end{keyword}

\hspace{7cm} \textit{In memory of Besnik Pumo}

\section{Introduction}

The contribution of Denis Bosq to functional data analysis and modeling is major for several reasons. First of all his work has the flavor of pioneering steps. The article \cite{Bosq1991} dates back to the very beginning of functional data. Of course some earlier papers investigate functions as statistical observations such as \cite{Dauxois1982}, \cite{Kleffe1973} but these authors usually confine themselves to infer without modeling, restricting to first and second order moment estimation or to correlation analysis. The second interesting point comes from the model itself : the ARH(1) as defined in \cite{Bosq1991} and soon studied by late Besnik Pumo and Tahar Mourid in their PhD thesis. It is seemingly the first model acting on functional data. The linear regression model appears only a few years later in \cite{CardotFLM99}. The ARH(1) will pave the way towards a specific domain of FDA which reveals fruitful: functional time series and processes. The reader interested with this topic is referred to \cite{hormann2010, Aue2015} and references therein for instance. At last the works by Denis Bosq  had a clear methodological impact by introducing tools from fundamental mathematics and connecting statistics with functional analysis and operator theory.

The ARH model has been widely investigated and generalized in several directions. The underlying Hilbert space was replaced by a space of continuous function in \cite{Pumo1998} then generalized to Banach spaces in \cite{Mourid2002}. The autoregressive operator was extended to linear processes with early results in \cite{Merlevde1996}. The celebrated monograph \cite{Bosq2000} sums up the main results on the topic (see also later \cite{Bosq2007}). At last some authors proposed variants of the original ARH by including derivatives (see the ARHD model \cite{Mas2009}) or by adding exogenous variable (\cite{Damon02} and their ARHX model).
 
The outline of the paper is the following. The ARH(1) model is introduced in the next section as well as the classical estimation procedure. Then we summarize Long Short Term Memory blocks that were selected for a numerical comparison. The results of our experiments are given in the last section. We first treated a large simulated dataset, then compared two temperature datasets and finally focused on a synthetic non-linear process. 

\section{The framework and the model}

Let $\mathbf{H}$ be a real separable Hilbert space endowed with an inner product $\left\langle .,.\right\rangle$
and a norm derived from it denoted $\left\Vert .\right\Vert $. In the rest of the paper the space $\mathbf{H}$ is the function space $L^2\left(\Omega\right)$, where $\Omega$ is assumed to be a real compact interval, usually $\left[0,T\right]$ for $T>0$. The space $\mathbf{H}$ could as well be of $W^{m,2}\left(\Omega\right)$ a Sobolev space with regularity index $m$.
\[
W^{m,2}=\left\{  f\in L^{2}\left(  \Omega\right)  : f^{\left(m\right)}\in L^{2}\left(
\Omega\right)\right\} .
\]

We will consider in the sequel a sample $\left(  X_{1},...,X_{n}\right)  \in\mathbf{H}^{n}$. When $X_{1}$ is
of functional nature its whole path is assumed to be observed. The expectation $\mathbb{E}X$ is a vector of $\mathbf{H}$ whenever it exists. The covariance operator of $X$ is denoted $\Gamma$. It is a positive, symmetric linear operator from $\mathbf{H}$ to $\mathbf{H}$ defined by :
\[
\Gamma=\mathbb{E}\left[\left(X-\mathbb{E}X\right)\otimes \left(X-\mathbb{E}X\right)\right]
\]
where $u \otimes v= \left\langle u, \right\rangle v$  is the tensor product notation for rank-one operators. The operator $\Gamma$ is trace-class and self-adjoint whenever $\mathbb{E}\left\|X\right\|^2 < +\infty$.
The centered autoregressive Hilbertian model reads :
\begin{equation}\label{arh}
X_{n+1}=\rho\left(X_n\right)+\varepsilon_{n+1},\qquad n \in \mathbb{Z}
\end{equation}
where $\left(\varepsilon_n\right)_{n \in \mathbb{N}}$ is a Hilbertian strong white noise and $\rho$ is a bounded linear operator acting from $\mathbf{H}$ to $\mathbf{H}$.
The model is studied in detail in \cite{Bosq2000}. Let $\left\|\rho\right\|_{\mathcal{L}}$ denote the classical -uniform- operator norm of $\rho$. We remind the reader this basic but crucial fact (see ibidem Theorem 3.1  p 74) : if $\left\|\rho\right\|_{\mathcal{L}}<1$ then the process $X_n$ solution of (\ref{arh}) is uniquely defined and stationary.
In the sequel we assume that $\left(X_n\right)_{n \in\mathbb{Z}}$ is both stationary and centered.

Estimation of $\rho$ is a difficult problem. Due to the functional framework, likelihood approaches are untractable in a truly infinite-dimensional framework. It can be shown that $\rho$ is the solution of a specific inverse problem. Namely if $D=\mathbb{E}\left[\left\langle X_n,\cdot\right\rangle X_{n+1}\right]$ is the cross-covariance of order 1 of the process :
\begin{equation} \label{eqnorm}
D=\rho \cdot \Gamma.
\end{equation}
The trouble with the above equation is that $\Gamma^{-1}$ does not exist unless $\Gamma$ is one-to-one. Then it is an unbounded linear operator, though measurable, and is defined on a domain $\mathcal{D} \varsubsetneq \mathbf{H}$. This domain is Borel-measurable but neither open nor closed and dense in $\mathbf{H}$. As a consequence deriving from (\ref{eqnorm}) $\rho=D \cdot \Gamma^{-1}$ is not correct since $\rho$ is defined on the whole $\mathbf{H}$ whereas $\Gamma^{-1}$ is not.

Any reasonable estimation procedures should simultaneously estimate $D$ and $\Gamma$ and regularize the latter in order to define say ``$\hat{\Gamma}^{\dag}$", approximation of $\Gamma^{-1}$. The estimation of $D$  and $\Gamma$ is usually simple though their empirical version :
\[
\hat{\Gamma}_n= \frac{1}{n} \sum_{i=1}^{n-1} X_i \otimes X_i \qquad \hat{D}_n= \frac{1}{n-1} \sum_{i=1}^{n-1} X_i \otimes X_{i+1}.
\]
At this point note that two smoothing strategies may be applied to stabilize the previous estimates : either smoothing the data (e.g. spline smoothing or decomposition in a basis of smooth function space) or smoothing the covariance operators only.

Approximation of $\Gamma^{-1}$ is usually more tricky and requires the computation of a regularized inverse denoted $\hat{\Gamma}^{\dag}$ above. This may be done directly by methods that are classical in inverse problem solving. For instance a ridge estimate provides then $\hat{\Gamma}_n^{\dag}=\left(\Gamma_n + T_n\right)^{-1}$ where $T_n$ is a regularizing (Tikhonov) matrix usually taken as $\alpha_n \mathbf{I}$ where $\alpha_n > 0$, $\alpha_n \downarrow 0$ and $\mathbf{I}$ denotes the identity matrix. Spectral (PCA based) regularization involve the random eigenelements of $\Gamma_n$, denoted $\left(\lambda_{i,n},\phi_{i,n}\right)\in \mathbb{R}^{+}\times \mathbf{H}$ where $\Gamma_n\phi_{i,n}= \lambda_{i,n} \phi_{i,n}$. A classical output is then :
\[
\hat{\Gamma}_n^{\dag}= \sum_{i=1}^{k_n} \frac{1}{\lambda_{i,n}}\phi_{i,n}\otimes\phi_{i,n}.
\]
where $k_n$ must be selected accurately.

Following again (\ref{eqnorm}) an estimate of $\rho $ then writes :
\[
\hat{\rho}_n= \hat{D}_n \cdot \hat{\Gamma}_n^{\dag} 
\]

The predictor is $\hat{\rho}_n\left(X_{n+1}\right)$ and stems from the preceding equation.  Note that the evaluation of $\hat{\rho}_n$ at $X_{n+1}$ simplifies the object under concern (the predictor is in $\mathbf{H}$ whereas $\rho$ is an operator on $\mathbf{H}$) and has a smoothing effect on the inverse problem mentioned above. Other results and further details may be found in \cite{Bosq2000,Mas2007}.  

\section{Long Short Term Memory Networks in a nutshell}

The question of predicting time series from neural networks is absolutely not new, see \cite{Bengio95}. When addressing the specific issue of prediction in time series, especially functional time series, Recurrent Neural Networks (RNN) appear as a natural and potentially effective solution. The basic RNNs architecture links a sequential input $X_n$ -typically with a stochastic dependence between $X_n$ and its past-  with an output $Y_n$ through an hidden layer $H_n$. The sequence $H_n$ is often compared with the hidden state in Hidden Markov chain modeling.  We refer to the beginning of \cite{Li2018} for a nice presentation of RNN's. The system is driven by the two following equations :
\begin{equation}\label{rnn}
\begin{aligned}
\left\{
\begin{array}{ll}
H_n= \sigma_h\left(\mathbf{A}X_n+\mathbf{B}H_{n-1}\right)  \\
Y_n= \sigma_y \left(\mathbf{C}H_{n}\right)
\end{array}
\right.
\end{aligned}
\end{equation}

where $\mathbf{A}$, $\mathbf{B}$ and $\mathbf{C}$ are matrices and $\sigma_h$ and $\sigma_y$ are two sigmoidal activation functions. Note that the previous matrices are fixed and not updated in the learning step. This specific structure enables the hidden layer to keep a memory of the past. As a consequence RNNs were successfully applied in speech recognition and more generally in treating dependent data indexed by time. Numerous variants of the RNN were proposed, many of them trying to make the network deeper, see \cite{Pascanu2014}.

\begin{figure}  
\centering                              
\includegraphics[scale=.5]{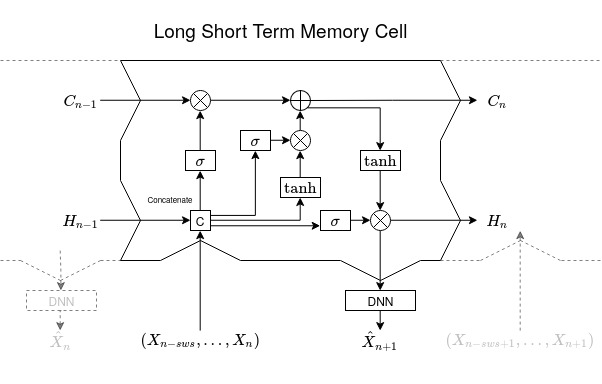}
\caption{Architecture of a LSTM block/unit}
\label{lstm}
\end{figure}

One of the most efficient variants of RNN are Long Short Term Memory units, trying to overcome the relative unability of RNN to capture long term dependence. They were introduced in the late 90's in \cite{Hochreiter97}.
Several tutorials may be found on the internet about LSTM. We give a sketch of the way LSTMs run but we refer the reader to \cite{Greff2017} for a formal description.
A key improvement in LSTMs over RNN relies on the addition of a cell state to the hidden space $H_n$ that appears in (\ref{rnn}). Figure \ref{lstm} shows the architecture of the single block LSTM which was used in this work. The cell state for unit $n$ is a vector denoted $C_n$. The cell state and the hidden state influence each other through three channels, also called gates. Roughly speaking $C_n$ updates $H_n$ within the LSTM block and will keep along the different layers the truly important information.
The three gates may be described in a few words. A first ``Forget" gate sweeps off the unimportant coordinates in the new input and in the current hidden state. Then the ``Input" gate aims at updating the cell state from $C_n$ to $C_{n+1}$. It applies a filter similar to the Forget gate on the concatenated vector $\left(H_{n-1},X_n\right)$. In parallel a $\tanh$ activation function is applied to the same vector, exactly like a single layer neural network. Then an Hadamard-product (coordinate-wise multiplication) merges the two preceding vectors. The by-product is added to the cell state posterior to the Forget Gate. The last step is the ``Output Gate" that first scales the current $C_n$ then filters $\left(H_{n-1},X_n\right)$ through a last sigmoid function. The resulting two vectors are Hadamard-multiplied, simultaneously generating the output and updating $H_{n-1}$ to $H_{n}$.

LSTMs gave rise to numerous variants. For instance some connections may be added between the three gates mentioned above and the current value of the cell state (referred to as ``peephole" connections). Conversely the LSTM architecture may be simplified like in the Gated Recurrent Unit (\cite{ChoMGBBSB14}).

\section{Numerical Experiments}

Below we consider only the one-step (functional) predictor : $\hat{\rho}_n\left(X_{n+1}\right)$. Keep in mind that this predictor provides a forecast of the whole path of the functional data on its period typically. All this comes finally down to a multi-step prediction in terms of univariate time-series.

After testing different stategies to evaluate our numerical results we adopted the following methodology. First we decided not to consider the rough Mean Square Error since it depends on the data's range and does not allow a comparison between datasets and methods. The MSE may also be hard to interpret. The Mean Absolute Relative Error (MARE) will be defined this way in our framework (we are aware that concatenating Absolute and Relative in the same acronym is not especially elegant) :
\[
\mathrm{MARE}=\frac{1}{n_{te}} \sum_{i=1}^{n_{te}}\mathrm{Abs}\left(X_i,\hat{X}_i\right)
\]
where $\hat{X}_i=\hat{\rho}_{n_{tr}}\left(X_i \right)$, $n_{te}$ is the size of the testing set and : 
\[
\mathrm{Abs}\left(X_i,\hat{X}_i\right)=\sum_{j=1}^{T}\frac{\left|X_i\left(t_j\right)-\hat{X}_i\left(t_j\right)\right|}{\left|X_i\left(t_j\right)\right|}.
\]
The integer $T$ is the (time-)grid size and the $t_j$'s are the discretization times . One of the problems with the above definition is that the denumerator may be null or very small. In order to avoid this problem all datasets were normalized to $\left[0.01,1\right]$.
Others normalizations were tested without any clear impact on the stability of the results. Notice that the absolute value at the denumerator in the definition of MARE may be removed.
Even if this methodology is certainly not optimal, it allows to compare -roughly at least- all the forthcoming results and to assess the overall performances of the methods.

\subsection{Simulations}\label{sect.simul}
The simulations were carried out with the \texttt{freqdom.fda} package, 
%(\cite{freqdom}).
We had the opportunity to process a rather large dataset this way. The data were generated by the \texttt{fts.rar} function. They follow a centered ARH(1) process with Gaussian white noise in a Fourier basis of dimension $2D+1$. This means that each data $X_i$ obeys (\ref{arh}) and is developed as a series made of a constant function and $D$ harmonics such as :
\[
X_i\left(t\right)=a_0+\sum_{k=1}^{D}\left\{a_k^{\left(i\right)}\cos\left(2\pi kt\right)+b_k^{\left(i\right)}\sin\left(2\pi kt\right)\right\}
\]

where $a_k^{\left(i\right)}$ and $b_k^{\left(i\right)}$ are sequences of real random variables. In order to ensure stationarity 50 burning iterations of the processes were conducted. The scenarii for the simulations depend on :
\begin{itemize}
\item Three different values for the Fourier basis dimension $D$,
\item Two different autocorrelation operators $\rho$ described just below.
\end{itemize}

The default autocorrelation operator of \texttt{fts.rar} was used. It is a large dimensional matrix whose row $i$-column $j$ cell $\rho_{i,j}$ is proportional to $exp\left(\left|i-j\right|\right)$ hence rapidly decreasing out of the diagonal. We also investigated the situation when the cells decrease more slowly : $\rho_{i,j} \propto \frac{1}{1+\left|i-j\right|^2}$. In both cases the Hilbert-Schmidt norm of $\rho$ was fixed so that $\left\|\rho\right\|_{HS}=0.5$.

\begin{figure}
\centering                                
\includegraphics[width=12cm,height=8cm]{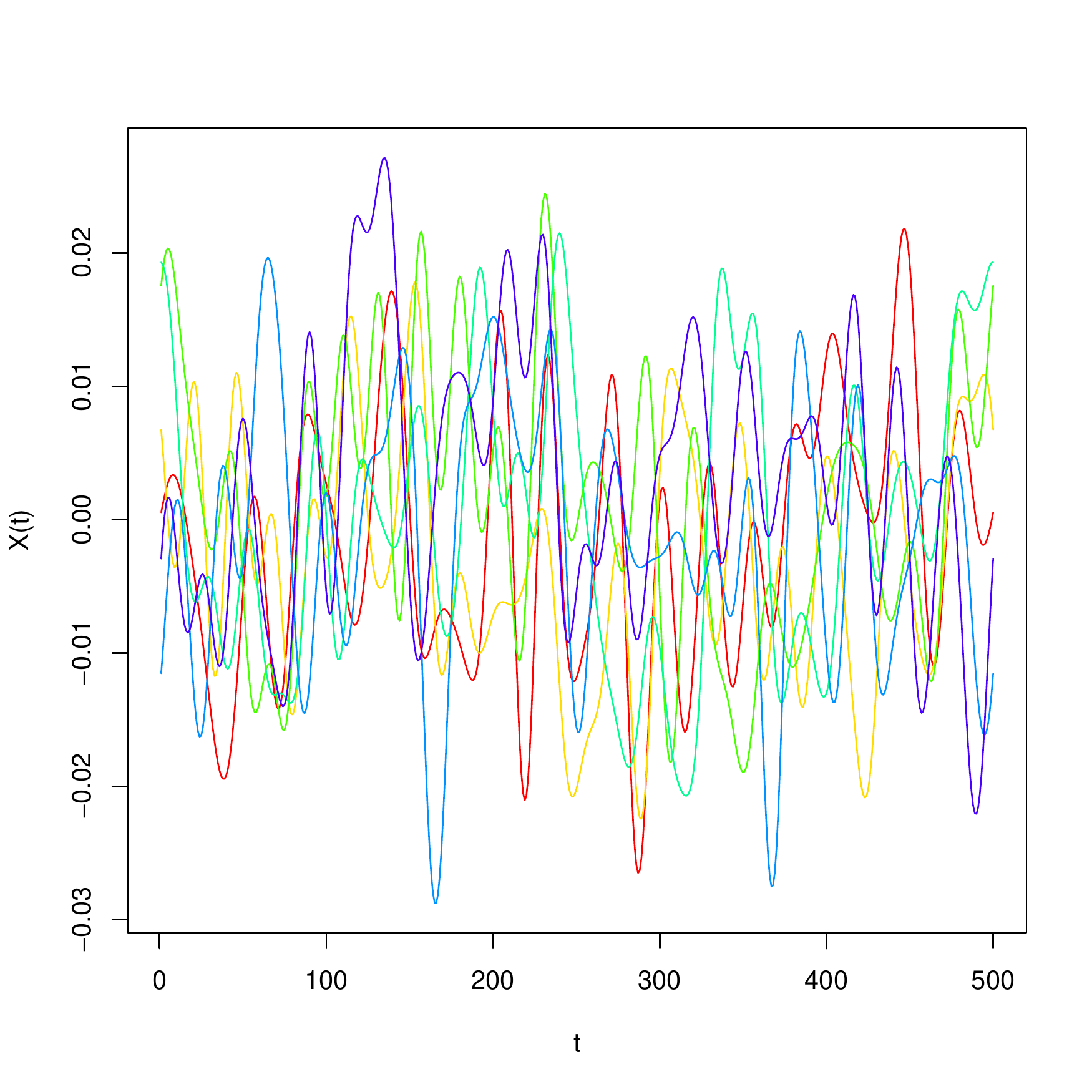}
\caption{Sample of size 5 of the simulated dataset with $D=21$ and $\rho$ of exponential type}
\label{simul-plot}
\end{figure}

Once generated in the basis the data were evaluated on a regular grid of size 500 in order to draw them and to compute their norms. The sample size was 1000. The data are consequently collected in a $\left(500\times1000\right)$ matrix. An overview of the data is given at Figure \ref{simul-plot}.

For both methodologies the initial dataset was first split in three subsets of size $n_{tr}=600$, $n_v=200$ and $n_{te}=200$ respectively for training, validation and test.
In the classical approach, the data matrix was processed as an fdata object by the \texttt{far} function of the package \texttt{far} by S. Guillas and J. Damon (\cite{far}). The previous package carries out the estimation of $\rho$ by the spectral cut (PCA) methodology and the prediction. The cross-validation step correctly determines the optimal value of $k_n$, around $2D+1$ in all situations.

The Neural Network part was conducted under Keras (see \cite{chollet2015keras}) with TensorFlow 2.0 as backend. The LSTM unit is followed by a dense layer whose output size equals the grid size (here 500). The learning rate for gradient descent was fixed to 1e-4. The training step is stopped when the MARE does not decrease anymore on 5 consecutive epochs on the validation set. The best epoch is then used for the testing step. The LSTM was carried out by taking into account the data in a sliding window of varying size (denoted SWS below). Here since the data are simulated according to an ARH(1) this optimal SWS is 1. In the case of an ARH(p) it would be obviously p. 

 Table \ref{mse.simul} displays the MARE values. We notice first that the autoregression operator structure (exponential or power 2) does not seem to have a clear impact. The MARE generally decreases when the latent dimension $D$ increases. A penalization term should certainly be added to balance this side-effect. Remind however that our goal here is to compare two methodologies. The ARH model was always optimally calibrated and provides the best results which is not surprising. We checked that the MARE decreases logically when the noise level in the model shrinks. Conversely the LSTM cell was not specifically designed for this data. The gap is not wide and seems rather promising in view of application on real data.

%% Please add the following required packages to your document preamble:
% \usepackage{multirow}

\begin{table}[] \label{mse.simul}
\centering
\begin{tabular}{|c||c||c||c|c|}
\hline
\multicolumn{1}{|l|}{$\rho$ type} & \multicolumn{1}{l|}{Effective   Dimension (2D+1)} & \multicolumn{1}{l|}{Stat pred MARE} & SWS & LSTM MARE \\ \hline
\multirow{6}{*}{exp}           & \multirow{2}{*}{21}                            & \multirow{2}{*}{\textbf{0.156}}     & 1   & 0.211     \\ \cline{4-5} 
                               &                                                &                                     & 2   & 0.209     \\ \cline{2-5} 
                               & \multirow{2}{*}{51}                            & \multirow{2}{*}{\textbf{0.140}}     & 1   & 0.193     \\ \cline{4-5} 
                               &                                                &                                     & 2   & 0.195     \\ \cline{2-5} 
                               & \multirow{2}{*}{81}                            & \multirow{2}{*}{\textbf{0.131}}     & 1   & 0.178     \\ \cline{4-5} 
                               &                                                &                                     & 2   & 0.177     \\ \hline
\multirow{6}{*}{pow}           & \multirow{2}{*}{21}                            & \multirow{2}{*}{\textbf{0.156}}     & 1   & 0.200     \\ \cline{4-5} 
                               &                                                &                                     & 2   & 0.202     \\ \cline{2-5} 
                               & \multirow{2}{*}{51}                            & \multirow{2}{*}{\textbf{0.141}}     & 1   & 0.178     \\ \cline{4-5} 
                               &                                                &                                     & 2   & 0.178     \\ \cline{2-5} 
                               & \multirow{2}{*}{81}                            & \multirow{2}{*}{\textbf{0.132}}     & 1   & 0.192     \\ \cline{4-5} 
                               &                                                &                                     & 2   & 0.191     \\ \hline
\end{tabular}
\caption{Simulated ARH : MARE for the statistical predictor versus LSTM}
\end{table}

\subsection{Real data}
\subsubsection{El Nino}
\begin{figure}  
\centering                              
\includegraphics[width=12cm,height=8cm]{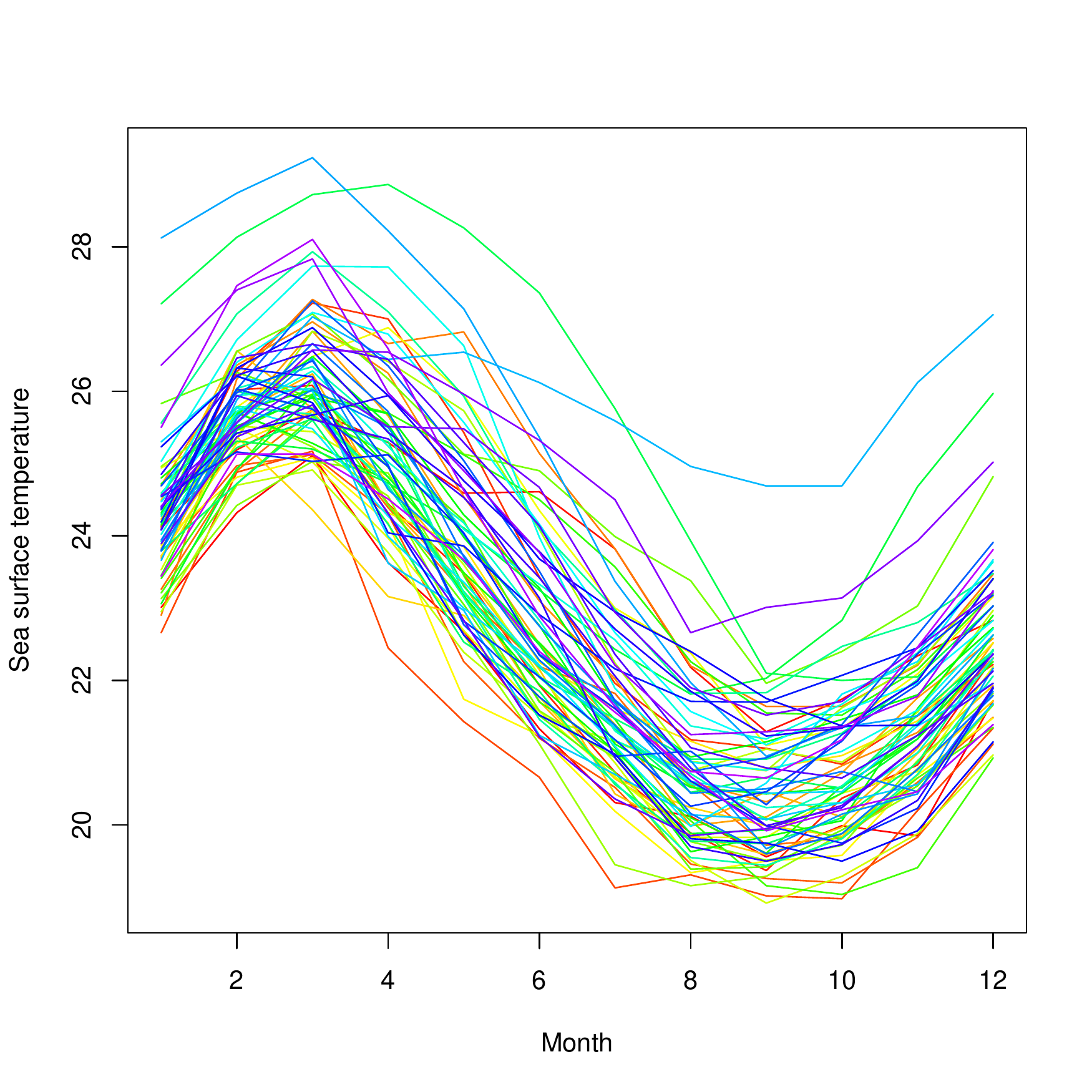}
\caption{The Sea Surface Temperature in El Nino dataset}
\label{elniplot}
\end{figure}

The El Nino dataset is one of the first which was studied in the framework of dependent functional data (see e.g. \cite{besse2000}). Our version comes from the \texttt{rainbow} package in R. It provides the sea surface temperature from January 1950 to December 2018 observed monthly. The bunch of curves is plotted at Figure \ref{elniplot}.

Out of 69 curves-data, 40 were used for training, 15 for validation and 14 for test. The modest size of the dataset restricted our study to SWS of size 1 and 2 only. The summary of MARE is given in Table \ref{elnino-mse1}.
\begin{table}[h] \label{elnino-mse1}
\centering
\begin{tabular}{|l||l|l|}
\hline
Stat. Predict MARE         & SWS & LSTM MARE \\ \hline
\multirow{2}{*}{0.226} & 1   & 0.301 \\ \cline{2-3} 
                          & 2   & 0.308 \\ \hline
\end{tabular}
\caption{El Nino Dataset : MARE for the statistical predictor versus LSTM}
\end{table}

The LSTM is again outperformed by the statistical predictor, but the MARE range, above 20\% is not good. At this point we must mention that we were faced with two main numerical issues concerning this meteorological dataset.

\begin{figure}     
\centering                           
\includegraphics[width=12cm,height=8cm]{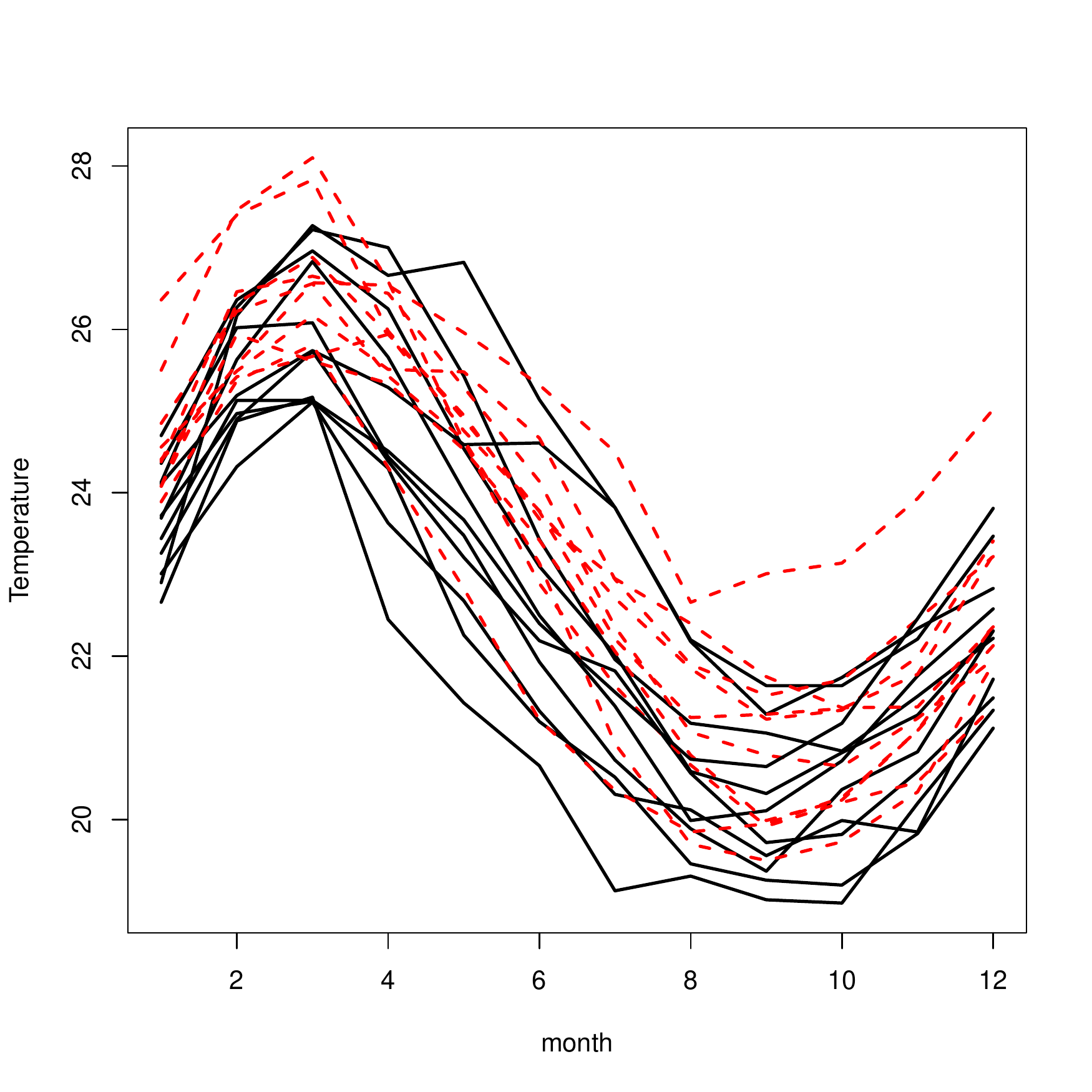}
\caption{The ten first (black solid) and ten last(red dashed) curves-data in the El Nino dataset}
\label{warm}
\end{figure}

First of all, even if we do not aim here at proving (again) the global warming, it seems that this fact could be retrieved from observations of the ten first versus the ten last curves-data as plotted on Figure \ref{warm}. The ten first are black-solid, the ten last are red-dashed. It is plain that sea temperature for the six first months of observations tend to be higher for recent years. As a consequence the basic assumption on stationarity of the data is not clearly fulfilled.

Second we need to underline the problems encountered when applying the usual strategy based on training, validation and testing for such dependent functional data. As explained earlier the training test is separated from the testing set by a validation interval containing 15 years of data. This strategy is clearly more sensitive to potential non-stationarity or slight perturbation in the model than in the situation where the sample is i.i.d. It results in a potential overfitting.
Ideally, validation, training and test should be performed continuously along the sample. But the model is not adapted to such strategies. Even if these results are not given here we noticed a substantial improvement of the MARE when using only a training and a testing set (without folding) plus a simple grid-search on $k_n$.

%\begin{thm}
%All conjectures are interesting, but some conjectures are more
%interesting than others.
%\end{thm}

%\begin{proof}
%Obvious.
%\end{proof}

\subsection{Bale temperature dataset}

\begin{figure}     
\centering                           
\includegraphics[width=12cm,height=8cm]{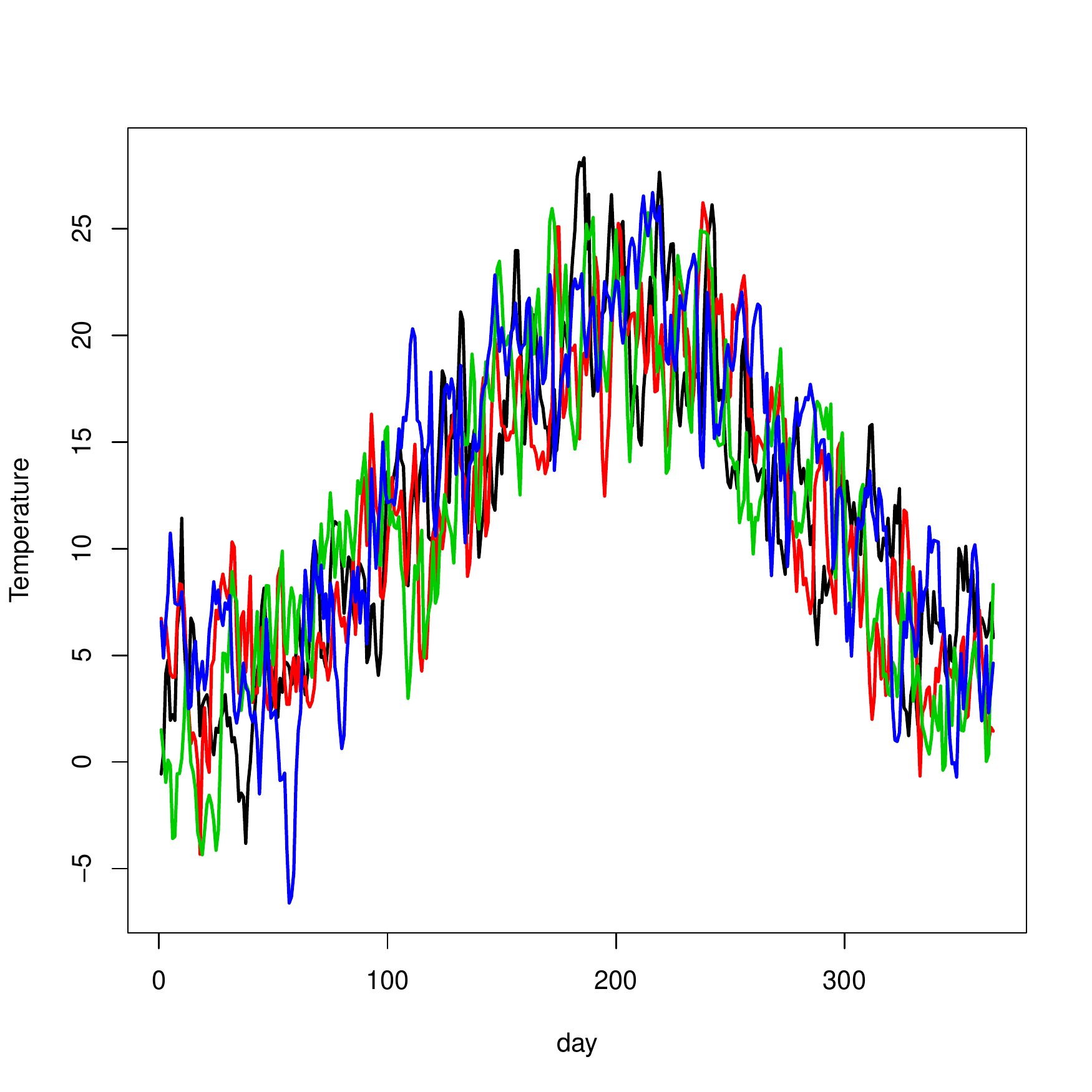}
\caption{A sample of 4 curves from the Bale temperature dataset. All data were picked in the testing set.}
\label{bale-temp}
\end{figure}

It may be interesting to compare the previous popular El Nino file with another temperature dataset retrieved freely from the website \url{https://www.meteoblue.com/fr/historyplus} and ranging from  $1985/1/1$ to  $2020/12/31$. The temperatures are recorded hourly in the city of Bale, Switzerland. We decided to consider the daily agregated data (the daily mean was used) in order to reduce drastically the ratio between the ambient dimension and the sample size. The data matrix is ($35\times365$) because all February 29th records were removed. The reader must notice that the sample size here is $n=35$ hence the half of El Nino's but the time frequency is the day (against the month).  A sample of curves is given in Figure \ref{bale-temp}.

The learning strategy was similar to El Nino. The prediction error is provided in Table \ref{bale-mse1}. Learning and calibration is performed on curves 1 to 30 and prediction on curves 31 to 35. The optimal dimension choice for the ARH predictor is  $k_n=3$. Conversely to El Nino the high sampling frequency of data allowed to explore SWS from 1 to 5. It is noticeable that the MARE are between 10\% and 15\% and improved with respect to El Nino. The statistical predictor is again slightly better than LSTM. All this tends to prove that the ARH model seems really competitive for these temperature datasets.

\begin{table}[h] \label{bale-mse1}
\centering
\begin{tabular}{|c||c|c|}
\hline
Stat. Predict MARE         & SWS & LSTM MARE \\ \hline
\multirow{5}{*}{0.116} & 1   & 0.245 \\ %\cline{2-5} 
                          & 2   & 0.133 \\ 
													& 3   & 0.126 \\ 
													& 4   & 0.140 \\ 
													& 5   & 0.125 \\ \hline
\end{tabular}
\caption{Bale temperature Dataset : MARE for the statistical predictor versus LSTM}
\end{table}

\subsection{Nonlinear ARH} \label{sect.non.linear}

\begin{figure}   
\centering                             
\includegraphics[width=12cm,height=8cm]{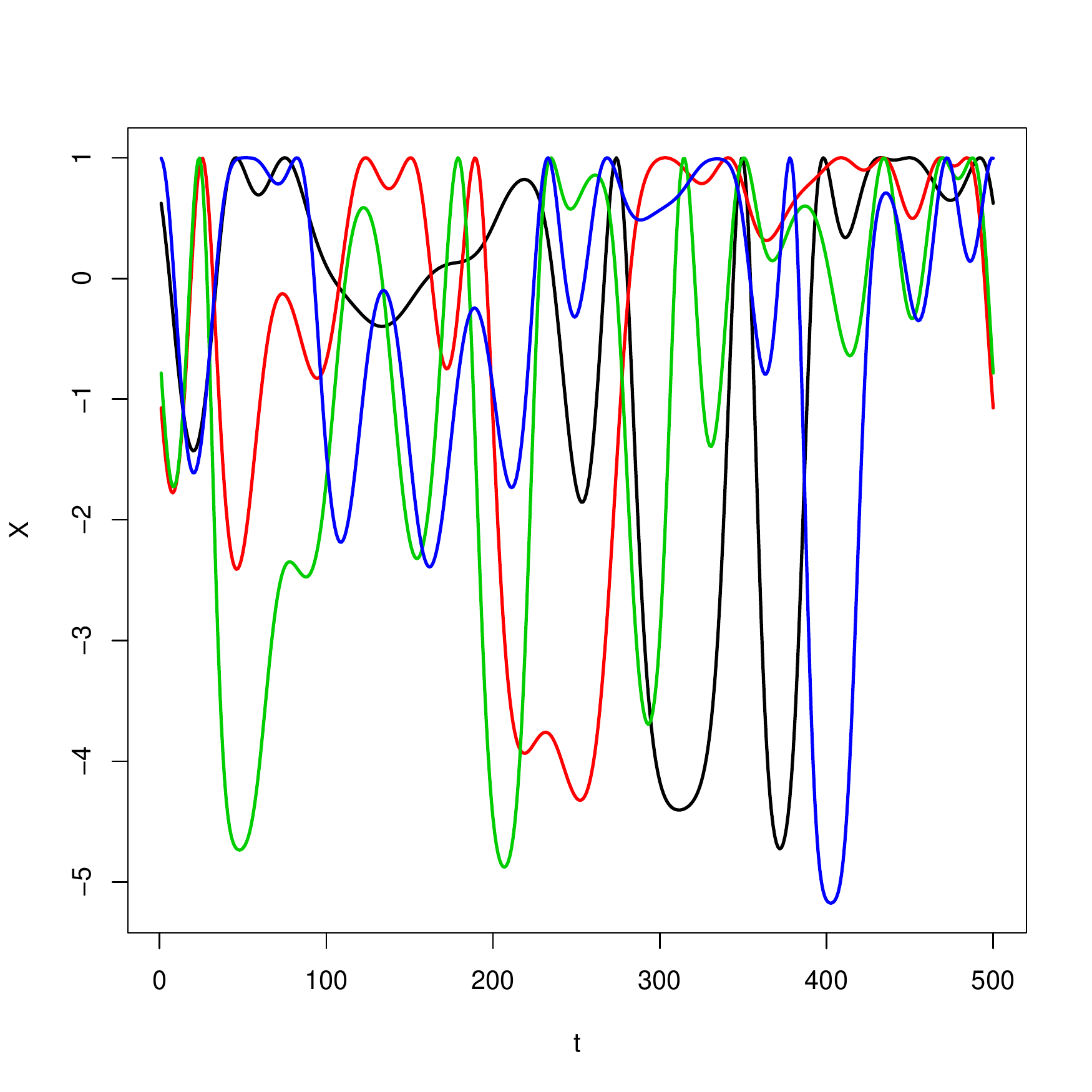}
\caption{Plot of $X_{200:203}^{n.l}$ for an overview of the nonlinear ARH process simulated on a grid of 500 time points. The whole sample size is 1000.}
\label{nonlin}
\end{figure}

Following the remark of a referee we investigated a situation which is less favorable to the ARH predictor and simulated a basic nonlinear functional autoregressive process. Start from a basic ARH equation simulating $X_n=\rho_{0}\left(X_{n-1}\right)+\epsilon_n$. Then construct the nonlinear process the following way :
\[
X_{n}^{n.l}\left(t\right)=3cos\left(10\pi\cdot X_n\left(t\right)\right)-2exp\left(-X_n\left(t\right)\right)
\]
A sample of four successive curves is plotted on Figure \ref{nonlin}. For a fair comparison with previous results, the dimension and sample size are the same as in section \ref{sect.simul}, respectively 500 and 1000.

\begin{table}[h] \label{nonlin-mse1}
\centering
\begin{tabular}{|c||c|c|}
\hline
Stat. Predict MARE         & SWS & LSTM MARE \\ \hline
\multirow{3}{*}{1.235} & 1   & 0.647 \\ %\cline{2-5} 
                          & 2   & 0.661 \\ 
													& 5   & 0.655 \\ \hline
\end{tabular}
\caption{Nonlinear autoregressive process : MARE for the statistical predictor versus LSTM}
\end{table}

The highly non-linear behaviour of $X_{n}^{n.l}$ is confirmed by the results in Table \ref{nonlin-mse1}. A ``quick and dirty" search gives an optimal $k_n$ around 70. The MARE are very high for this synthetic dataset close to a white noise. Anyway, despite this fact, the LSTM performs almost twice better than the statistical predictor. 
 
\section{Conclusion}
This work attempts to compare the historical/statistical track for prediction in ARH models with a Neural Network approach centered on LSTMs. Data and code are available at \url{https://gitlab.com/arh-lstm/}. Several facts should be underlined in order to show the limits of our results.
\begin{itemize}
\item We did not study here the impact of the sampling frequency i.e. the size of discretization grid for the functional data. We noticed however some improvement between the El Nino and the Bale dataset. On this basis nothing solid should be stated however. We could have also focused on the effect of the sample size or of the $\rho$ operator norm on the accuracy of the results.
\item The architecture used here is simplistic because based on a single LSTM block. Introducing some depth by adding several layers of LSTM should certainly improve the predictions of the simulated dataset. El Nino is certainly not suited to a sequence of cells.
\item We used the discretized version of the functional data coming down to a large dimensional input vector (up to size 500 here). Clearly feeding the network with the Fourier coefficient instead leads to a more compact entry and paves the way to another approach.  
\end{itemize}

Our framework was centered on the functional autoregressive process of order 1 and may be restrictive in some way. The design of LSTM is general enough to foster a wider investigation : autoregressive processes of order $p>1$ or even more general functional times series with linear or non-linear dependence structure. Further work is in progress in order to compare the numerical performance of Neural Networks strategy against functional non-parametric techniques such as kernel-regression in this setting of dependent functional data.

\bibliographystyle{apa} 
\bibliography{mybib}

\end{document}